\documentclass[aps,pra,twocolumn,groupedaddress,showpacs,floatfix]{revtex4}
\usepackage{graphicx}
\usepackage{amsmath}
\usepackage[dvips]{color}
\def\ket#1{|#1\rangle}
\def\bra#1{\langle#1|}

\begin{document}

\title{Optical Control of the Scattering Length and Effective Range\\ for Magnetically Tunable Feshbach Resonances in Ultracold Gases}

\author{Haibin Wu$^{1,2}$ and J. E. Thomas$^{1}$}
\affiliation{$^{1}$Department of  Physics, North Carolina State University, Raleigh, NC 27695, USA}
\affiliation{$^{2}$ State Key Laboratory of Precision Spectroscopy, Department of Physics, East China Normal University, Shanghai 200062, China}


\date{\today}

\begin{abstract}
We describe two-field optical techniques to control interactions in Feshbach resonances for two-body scattering in ultra-cold gases. These techniques  create a molecular dark state in the closed channel of a magnetically tunable Feshbach resonance,  greatly suppressing optical scattering compared to single optical field methods. The dark-state method enables control of the effective range,   by creating narrow features that modify  the energy dependence of the scattering phase shift, as well as control of the elastic and inelastic parts of the zero-energy s-wave scattering amplitude. We determine the scattering length and the effective range from an effective range expansion, by calculating the momentum-dependent scattering phase shift from the two-body scattering state.

\end{abstract}

\maketitle

\section{Introduction}

Ultracold atomic gases with controllable interactions are now widely studied by exploiting collisional (Feshbach) resonances~\cite{OHaraScience,RMP2008,ZwergerReview,ZwierleinFermiReview}, where the interaction strength is controlled by means of a bias magnetic field. In a  recent theoretical paper~\cite{WuOptControl}, we suggested a general ``dark-state" optical method  for widely controlling the scattering length near a magnetic Feshbach resonance, while suppressing spontaneous scattering by quantum interference. In this paper, we provide a more detailed treatment of the method and show that the molecular dark-state method  enables control of both  the scattering length and the effective range in the two-body collisions of ultra-cold gases.

Precision optical control of  the scattering length and effective range in two-body scattering enables rapid temporal control and high resolution spatial control of the interaction strength near a Feshbach resonance, opening many new fields of study, such as non-equilibrium strongly interacting Fermi gases~\cite{BulgacYoon09}. For example, the natural time scale in a Fermi gas is the ``Fermi-time,"  the time $\tau_F$ for an atom at the Fermi surface to move a de Broglie wavelength, i.e, $\tau_F=\lambda_F/v_F\simeq\hbar/E_F$. For Fermi energies $E_F$ in the $k_B\times 1\,\mu$K regime, $\tau_F$ is several $\mu$s. To explore non-equilibrium dynamics on this time scale~\cite{BulgacYoon09} requires fast control of interactions, which is readily achieved with optical methods. In simulating neutron matter, the unitary Fermi gas provides the simplest model, where the scattering length is large compared to the interparticle spacing. A more realistic model of neutron matter can be realized by adjusting both  the scattering amplitude and the effective range in a trapped Fermi gas, to achieve the known ratios for neutrons, where the effective range is comparable to the interparticle spacing~\cite{PethickEffRange}. Of great interest,  as discussed below, is that using optical control, the effective range can even be made large and {\it negative}. This is especially interesting for the narrow Feshbach resonance in $^6$Li, where recent theory~\cite{HoNarrowFB} suggests that a Fermi gas can be even more strongly interacting than for the broad resonance in the unitary regime.

 In contrast to Bose gases, which suffer from three-body inelastic processes near a resonance, two-component Fermi gas mixtures are stable as a result of the Pauli principle,  and can be rapidly cooled to quantum degeneracy by evaporation in the resonant regime~\cite{OHaraScience}. Typically, in a Feshbach resonance, an external magnetic field controls the interaction strength between spin-up and spin-down atoms, by tuning the  energy of an incoming, colliding atom pair  into resonance with that of a bound molecular state in an energetically closed channel~\cite{Stoof,ChinFeshbach}.
 Optical control of Feshbach resonances has been explored previously in Bose gases~\cite{GrimmFB,PLettOFB} and currently is receiving substantial attention~\cite{RempeOptControl}. Optical Feshbach resonances (OFR), which employ photoassociation light to drive a transition from the continuum of the incoming atom pair state to an excited molecular bound state, has been proposed and experimentally observed~\cite{WalravenOptTuning, JulienneOptTuning,PLettOFB,TheisOFB,GrimmOFB,enomoto}. However, light-induced inelastic collisions and the accompanying loss limit its practical applicability. Optical Feshbach resonances for control of higher partial waves, such as p-wave scattering of $^{171}$Yb,  has been suggested~\cite{Goyal-th-optical-p-wave} and demonstrated recently~\cite{Yamazaki-exp-optFB-p-wave}. OFR also has been studied by using a narrow intercombination line of  a bosonic gas $^{88}$Sr, with the laser frequency tuned far away from  resonance~\cite{YeOFB}. Submicron-scale spatial modulation of an inter-atomic interaction has been observed in an alkaline-earth atomic condensate~\cite{yamazaki}. Recently, Bauer and coworkers~\cite{RempeOptControl} have used a single optical field to control the scattering length near a magnetic Feshbach resonance  by driving a transition between the resonant ground state  and an excited molecular state in the closed channel. This elegant method enables substantial tuning of the scattering length, but a large laser intensity  and a large frequency detuning are required for suppressing the light-induced loss~\cite{RempeOptControl}. The use of electromagnetically induced transparency (EIT) to control Feshbach resonances was suggested by Harris~\cite{HarrisFB}. Deb~\cite{Deb} has suggested that quantum inference between photoassociation and magnetic Feshbach s-wave resonance amplitudes permits control of the scattering length and suppression of inelastic scattering. The method enables control of higher order partial waves as well. It is clear that the development of improved quantum interference methods for achieving wide tunability of scattering parameters while suppressing light-induced loss and heating~\cite{HarrisFB,Deb,WuOptControl} will greatly extend the applicability of optical control methods.

\begin{figure}[htb]
\begin{center}
\includegraphics[width=2.75in]{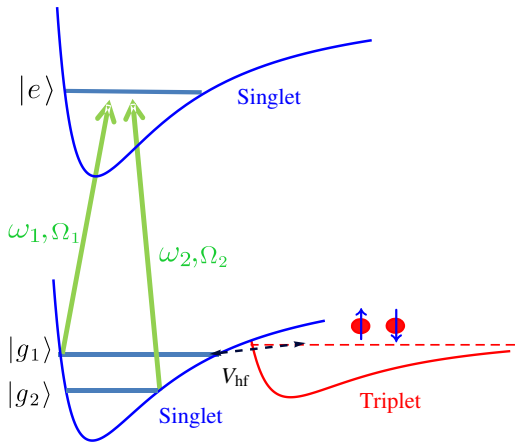}
\caption[example]
   { \label{fig:fig1}
(color online). Scheme for ``dark-state" optical control of a Feshbach resonance using two closed-channel molecular states. Optical fields of frequencies $\omega_{1}$ and $\omega_{2}$ and Rabi frequencies $\Omega_{1}$ and $\Omega_{2}$, respectively,  couple two singlet ground  molecular states $\ket {g_{1}}$ and $\ket {g_{2}}$ to the excited singlet molecular state $\ket e$;  $V_{HF}$ is the hyperfine  coupling  between the incoming atomic pair state in the open (triplet) channel and $\ket{g_{1}}$, which is responsible for a magnetically controlled Feshbach resonance. }
\end{center}
\end{figure}

In the paper, we analyze the molecular dark-state method proposed in Ref.~\cite{WuOptControl} to determine both the optically controlled s-wave scattering length and the effective range from the relative momentum dependence of the scattering phase shift. The basic scheme, Fig.~\ref{fig:fig1},  is illustrated for a pair of atoms in two hyperfine states (denoted spin-up and spin-down), which undergoes an s-wave collision in the ground electronic state triplet molecular potential (open channel). The hyperfine interaction couples the scattering continuum of the open channel to a bound singlet vibrational state $\ket {g_{1}}$ in the energetically closed channel. An applied bias magnetic field $B$ tunes the total energy of the colliding atom pair downward, near $\ket{g_1}$, producing a collisional (Feshbach) resonance. Two optical fields with frequencies $\omega_{1}$  and $\omega_{2}$ couple $\ket {g_{1}}$ and  $\ket {g_{2}}$ to the electronically excited singlet vibrational state $\ket e$, creating a ``dark" state.  All three  molecular levels in the closed channel are assumed to have the same total nuclear-electron-spin state as $\ket{g_1}$, so that both optical transitions to the level $\ket{e}$ are fully allowed. For example, $\ket {g_{2}}$ can be a different singlet vibrational state from $\ket {g_{1}}$. To determine the momentum dependence of the s-wave scattering phase shift in the presence of the light fields, we use a method similar to that employed by Fano~\cite{Fano}, to treat the coupling of  an open channel  continuum to a bound  state in an energetically closed channel.

  The primary results of this paper show how both the s-wave scattering length (eq.~\ref{eq:2.6}) and the effective range (eq.~\ref{eq:effrangeopt3}) can be controlled using two optical fields, which alter the momentum-dependent phase shift of the scattering state.   This method  is applicable to both broad and narrow Feshbach resonances, as occur in $^6$Li. In Appendix~\ref{simplemodel}, we present a simple model to determine the parameters for Feshbach resonances with large background scattering lengths and apply it to $^6$Li.\\

\section{Time-Dependent Scattering State}

We will consider first the relevant states for a broad Feshbach resonance. As a concrete example, we will use a mixture of the two lowest hyperfine states  of $^6$Li, denoted $\ket{1}$ and $\ket{2}$.  In a bias magnetic field, the atoms interact by s-wave scattering, and have a total energy determined by the incoming relative kinetic energy, the combined hyperfine energies, and the Zeeman energies.

For a pair of atoms, one each in states $\ket{1}$ and $\ket{2}$, the total magnetic quantum number, $M=0$, is conserved in a bias magnetic field, $B_z$. There are 5 two-atom states for $M=0$, which can be written in the ``interior" singlet-triplet basis $\ket{S,m_S;I,m_I}$. There are two singlet states $\ket{0,0; 0,0}$, $\ket{0,0;2,0}$, which differ in the total nuclear spin $I=0,2$, and are degenerate in the absence of hyperfine interactions. In addition, there are three triplet states
$\ket{1,-1;1,1}$, $\ket{1,0;1,0}$, and $\ket{1,1;1,-1}$. As shown below, however, only one singlet state and one triplet state are relevant. At high bias magnetic fields, the antisymmetric combination of states $\ket{1}$ and $\ket{2}$ is predominantly the triplet electronic spin state $\ket{1,-1;1,1}$.

A Feshbach resonance in the s-wave channel arises when the bias magnetic field tunes the total energy of a colliding atom pair in the open triplet channel into resonance with a bound molecular state in the energetically closed singlet channel. For $^6$Li, the Feshbach resonant state, denoted by $\ket{g_1}$, is the ground-singlet $38^{th}$ molecular vibrational state. As this singlet bound state is lower in energy than the incoming unbound triplet states at zero bias field, the triplet state $\ket{1,-1;1,1}$, which tunes downward with increasing B-field, is responsible for the resonance, which arises from the hyperfine coupling between the triplet and singlet channels. The other two triplet states  tune negligibly or tune upward, and can be neglected for the broad Feshbach resonance at 834 G.

Restricting attention to the  3 interior states, $\ket{0,0;0,0}$, $\ket{0,0;2,0}$, and $\ket{1,-1;1,1}$, we consider the origin of the broad and narrow Feshbach resonances in $^6$Li. The hyperfine interaction couples each of the singlet states to the triplet. However, one combination of $\ket{0,0;0,0}$ and $\ket{0,0;2,0}$  has a {\it zero} hyperfine matrix element with the triplet state $\ket{1,-1;1,1}$. This uncoupled singlet state is $ \ket{\psi_N}=(\ket{0,0;0,0}+2\sqrt{2}\ket{0,0;2,0})/3$ and
 is responsible for the {\it narrow} Feshbach resonance in $^6$Li, near 543 G~\cite{zerocross,HuletNarrow}. The narrow resonance arises from  a {\it second order} coupling of $\ket{\psi_N}$ to $\ket{1,-1;1,1}$ through the $\ket{1,0;1,0}$ triplet state, which is far detuned from the singlet states compared to the hyperfine energy.

To treat the broad Feshbach resonance, we can therefore consider just {\it two} states. The first is the combination of singlet states which is {\it orthogonal} to $\ket{\psi_N}$, $\ket{g_1}\equiv(2\sqrt{2}\,\ket{0,0;0,0}-\ket{0,0;2,0})/3$.
The second is the triplet state $\ket{T}\equiv \ket{1,-1;1,1}$, for which the diagonal element of the Zeeman-hyperfine energy is $E_T\equiv-a_{HF}/2-2\mu_B\,B$,
where $a_{HF}/h = 152.1$ MHz is the hyperfine coupling constant and $\mu_B$ is the Bohr magneton, $\mu_B/h\simeq 1.4$ MHz/G.

We take the unperturbed two-state Hamiltonian, in the absence of optical fields, to be
\begin{eqnarray}
H_0&=&E_{g_1}\ket{g_1,v_1}\bra{ g_1,v_1}+E_{g_2}\ket{g_2,v_2}\bra{g_2,v_2}\nonumber\\
&+&E_e\ket{e,v_e}\bra{e,v_e}+V_{HF}\left(|g_1\rangle\langle T|+|T\rangle\langle g_1|\right)\nonumber\\
&+&\left(E_T+\frac{p^2}{m}\right)\sum_{k_1}\ket{T,k_1}\bra{T,k_1}\nonumber\\
&+&(E_T + E_b)\ket{T,b}\bra{T,b},
\label{eq:Hamiltonian}
\end{eqnarray}
where $\ket{g_1,v_1},\,\ket{g_2,v_2}$ are two ground electronic-vibrational states in the singlet molecular potential and $\ket{e,v_e}$ is an excited singlet electronic-vibrational state. $\ket{T,k_1}$ is a  triplet continuum state. To model a {\it large positive} background scattering length $a_{bg}$, we include in the model a  near threshold triplet bound state $\ket{T,b}$, where $E_b=-\hbar^2/(ma_{bg}^2)$. We will omit this term for small background scattering lengths and for large negative background scattering lengths, but it will be important for our discussion of the effective range.  In eq.~\ref{eq:Hamiltonian}, the singlet-triplet hyperfine coupling  for the broad 1-2 resonance in $^6$Li is $V_{HF}=\langle g_1|H'_{HF}|T\rangle=-3\,a_{HF}/(2\sqrt{3})=-h\times 131.6\,{\rm MHz}$, where $H'_{HF}$ is the hyperfine interaction.

As the two optical fields used for this method can be tuned relatively close to resonance, the isolated resonance  model~\cite{JulienneOptTuning} is adequate, in contrast to single field methods employing very large detunings and high intensity to avoid optical scattering, where this approximation fails~\cite{YeOFB}.
According to Fig.~\ref{fig:fig1}, the two optical fields introduce a perturbation,
\begin{eqnarray}
H'&=&-\frac{\hbar\Omega_1}{2}\,e^{-i\omega_1 t}\ket{e,v_e}\bra{g_1,v_1}\nonumber\\
& &-\frac{\hbar\Omega_2}{2}\,e^{-i\omega_2 t}\ket{e,v_e}\bra{g_2,v_2}+h.c.,
\end{eqnarray}
where we have used the rotating-wave approximation. Here, $\Omega_i$ are the Rabi frequencies for the optical transitions, which include the vibrational Franck-Condon factors.

We take the time-dependent scattering state, for a triplet input state of energy $E=E_T+\hbar^2k^2/m$, to be of the form,
\begin{eqnarray}
&\hspace*{-0.1in}\ket{\Psi_E(t)}=c_1\,\ket{g_1,v_1}+c_2\,\ket{g_2,v_2}+c_e\,\ket{e,v_e}\nonumber\\
&\hspace*{-0.1in}+\,c_{T}(k)\,\ket{T,k}+c_{T}(b)\ket{T,b}+\sum_{k'\neq k}c_{T}(k')\,\ket{T,k'},
\label{eq:scattstate}
\end{eqnarray}
 where $\ket{T,k}$ is the {\it selected input} triplet scattering state (in the open channel), with $\hbar k$ the relative momentum of the colliding atoms. Note that all of the amplitudes, $c_1$, etc., are both $k$-dependent and time-dependent. The sum denotes the rest of the triplet continuum for $k'\neq k$, which leads to the {\it principal part} of the integral obtained in the continuum limit.  For $r\rightarrow\infty$, the triplet continuum background states (box normalization) have the asymptotic form
 \begin{equation}
 \langle r |k \rangle=\frac{1}{\sqrt{V}}\frac{\sin(kr+\delta_{bg})}{kr},
 \label{eq:backgroundstate}
 \end{equation}
where $\delta_{bg}$ denotes the phase shift arising from the effective triplet potential, which corresponds to the background scattering length $a_{bg}$, i.e., $\delta_{bg}\rightarrow -k\,a_{bg}$ for $k\rightarrow 0$.

To obtain the scattering solution, we solve the time-dependent  Schr\"{o}dinger equation,
\begin{equation}
(H_0+H')\ket{\Psi_E(t)}=i\hbar\frac{\partial}{\partial t}\ket{\Psi_E(t)}.
\label{eq:Schroedinger}
 \end{equation}
 Taking projections of
 eq.~\ref{eq:Schroedinger} with $\bra{g_1,v_1}$, we obtain
 \begin{eqnarray}
 \dot{c}_1&=&-\frac{i}{\hbar}E_{g_1}c_1-ig^*(k)\,c_T(k)\nonumber\\
 & &-\,i\sum_{k'\neq k}g^*(k)\,c_T(k')-ig^*_b\,c_T(b)\nonumber\\
 & &+\,i\,\frac{\Omega^*_1}{2}\,e^{-i\omega_1t}c_e.
 \label{eq:c1}
 \end{eqnarray}
 where we have defined the hyperfine coupling strengths
 \begin{eqnarray}
 \hbar g(k)&\equiv& V_{HF}\,\langle k|v_1\rangle\\
 \hbar g_b&\equiv&V_{HF}\,\langle b|v_1\rangle.
 \label{eq:coupling}
 \end{eqnarray}
 Here, the overlap integrals are determined from the spatial wavefunctions $\langle r|v_1\rangle$ and $\langle r|b\rangle$, which  are vibrational wavefunctions in the singlet and triplet channel, respectively, and $\langle r |k \rangle$ is given by eq.~\ref{eq:backgroundstate}.
Similarly, projecting onto $\bra{g_2,v_2}$  yields
\begin{equation}
\dot{c}_2=-\frac{i}{\hbar}E_{g_2}c_2+i\frac{\Omega_2^*}{2}\,e^{i\omega_2t}c_e,
\label{eq:c2}
\end{equation}
 while $\bra{e,v_e}$ gives for the excited singlet electronic state amplitude
 \begin{equation}
 \dot{c}_e=-i\frac{E_e}{\hbar}c_e-\frac{\gamma_e}{2}c_e+i\frac{\Omega_1}{2}\,e^{-i\omega_1t}c_1+
 i\frac{\Omega_2}{2}\,e^{-i\omega_2t}c_2.
 \label{eq:ce}
 \end{equation}
 Here, we have added a phenomenological decay term corresponding to the excited state molecular spontaneous emission rate $\gamma_e$. As the decay rate arises from a spontaneous emission into a broad range of ground singlet vibrational states, the arrival rate of molecules into the ground states $\ket{g_1}$ and $\ket{g_2}$ is negligible.
Finally, projecting onto  $\bra{T,k''}$ and $\bra{T,b}$ yield
\begin{equation}
\dot{c}_T(k'\neq k)=-\frac{i}{\hbar}\left(E_T+\frac{\hbar^2k'^2}{m}\right)c_T(k')-ig(k')\,c_1
\label{eq:cTk}
\end{equation}
and
\begin{equation}
\dot{c}_T(b)=-\frac{i}{\hbar}(E_T+E_b)\,c_T(b)-ig_b\,c_1,
\label{eq:cTb}
\end{equation}
where we note that the triplet bound state and triplet continuum states are orthogonal.

These time-dependent amplitude equations are easily solved.  We expect that all of the triplet amplitudes will have a time-dependent  phase factor $\exp(-iEt/\hbar)$, as the asymptotic input and output states are energy eigenstates in the triplet potential. Hence, we take $c_T(k')=b_T(k')\,\exp(-iEt/\hbar)$ for all $k'$ including $k$ and $c_T(b)=b_T(b)\,\exp(-iEt/\hbar)$. Then, eq.~\ref{eq:cTk}~and~\ref{eq:cTb} yield,
\begin{eqnarray}
\dot{b}_T(k')&=&-\frac{i}{\hbar}\left(E_T+\frac{\hbar^2k'^2}{m}-E\right)\,b_T(k')\nonumber\\
& &\hspace*{0.25in}-ig(k')c_1\,e^{iEt/\hbar}
\label{eq:bTk1}
\end{eqnarray}
for $k'\neq k$ and
\begin{eqnarray}
\dot{b}_T(b)&=&-\frac{i}{\hbar}(E_T+E_b-E)\,b_T(b)\nonumber\\
& &\hspace*{0.25in}-ig_b\,c_1\,e^{iEt/\hbar}.
\label{eq:bTb1}
\end{eqnarray}
It is obvious that we can  eliminate the explicit time dependent phase factors in eqs.~\ref{eq:bTk1}~and~\ref{eq:bTb1} by taking $c_1=b_1\,\exp(-iEt/\hbar)$. Hence,
\begin{equation}
\dot{b}_T(k')=-\frac{i}{\hbar}\left(E_T+\frac{\hbar^2k'^2}{m}-E\right)\,b_T(k')-ig(k')b_1
\label{eq:bTk}
\end{equation}
and
\begin{equation}
\dot{b}_T(b)=-\frac{i}{\hbar}(E_T+E_b-E)\,b_T(b)-ig_b\,b_1.
\label{eq:bTb}
\end{equation}
Using $c_1=b_1\,\exp(-iEt/\hbar)$ in eq.~\ref{eq:c1} we obtain an equation for $\dot{b}_1$. We eliminate the explicit time-dependent phase factor that appears in the $\dot{b}_1$ equation by taking $c_e=b_e\,\exp[-i(\omega_1+E/\hbar)t]$, which yields
\begin{eqnarray}
\dot{b}_1&=&-\frac{i}{\hbar}(E_{s_1}-E)\,b_1-ig^*(k)\,b_T(k)\nonumber\\
& &-i\sum_{k'\neq k}g^*(k')\,b_T(k')\nonumber\\
& &-ig^*_b\,b_T(b)+i\frac{\Omega_1^*}{2}b_e.
\label{eq:b1}
\end{eqnarray}
Using $c_e=b_e\,\exp[-i(\omega_1+E/\hbar)t]$ in eq.~\ref{eq:ce}, we obtain an equation for $\dot{b}_e$. We eliminate the explicit time-dependent phase factor in that equation by taking $c_2=b_2\,\exp[+i(\omega_2-\omega_1-E/\hbar)t]$. Then,
\begin{equation}
\dot{b}_e=i\Delta_e\,b_e-\frac{\gamma_e}{2}b_e+i\frac{\Omega_1}{2}b_1+i\frac{\Omega_2}{2}b_2,
\label{eq:be}
\end{equation}
where, the one-photon detuning $\Delta_e$ is given by
\begin{equation}
\Delta_e\equiv \omega_1-\frac{E_e-E}{\hbar}.
\label{eq:Deltae}
\end{equation}
Finally, using $c_2=b_2\,\exp[+i(\omega_2-\omega_1-E/\hbar)t]$ in eq.~\ref{eq:c2}, we obtain
\begin{equation}
\dot{b}_2=-i\delta\,b_2+ i\frac{\Omega_2^*}{2}b_e,
\label{eq:b2}
\end{equation}
where the two-photon detuning $\delta$ is given by
\begin{equation}
\delta \equiv \omega_2-\omega_1-\frac{E-E_{g_2}}{\hbar}.
\label{eq:delta}
\end{equation}

 For $^6$Li, where the molecular $\gamma_e\simeq 2\gamma_{spont}=2\pi\times 11.8$ MHz, and for other atoms with a large spontaneous decay rate, we can eliminate the excited state amplitude by making an adiabatic approximation, where we assume $\dot{b}_e<<\gamma_e\,b_e/2$. In this case, the excited state amplitude tracks $b_1$ and $b_2$.
 \begin{equation}
 b_e\simeq -\frac{\Omega_1\,b_1+\Omega_2\,b_2}{2(\Delta_e+i\gamma_e/2)}.
 \label{eq:beeq}
 \end{equation}

 We look for scattering state solutions of the amplitude equations where $\dot{b}_1,\dot{b}_2,\dot{b}_T(k)$ and $\dot{b}_T(b)$ are all zero. Then, eq.~\ref{eq:b2} yields $b_2=\Omega_2^*b_e/(2\delta)$, which with eq.~\ref{eq:beeq} yields $b_e$ in terms of $b_1$. Then eq.~\ref{eq:bTk} yields
 \begin{equation}
 b_T(k'\neq k)=\frac{\hbar g(k')\,b_1(k)}{E-E_T-\hbar^2k'^2/m}
 \label{eq:bTk'}
 \end{equation}
 and eq.~\ref{eq:bTb} gives $b_T(b)=\hbar g_b\,b_1(k)/(E-E_T-E_b)$. Using these results in eq.~\ref{eq:b1}, we obtain $b_1(k)$ in terms of the amplitude of the {\it input} triplet scattering state amplitude $b_T(k)$,
 \begin{equation}
 b_1(k)=\frac{\hbar\,g^*(k)\,b_T(k)}{D(E)},
 \label{eq:b1k}
 \end{equation}
 where
\begin{equation}
D(E)\equiv E-E_{g_1}-\Sigma_E(k)-\frac{\hbar|\Omega_1|^2}{4[\Delta_e+i\frac{\gamma_e}{2}+\frac{|\Omega_2|^2}{4\delta}]}.
\label{eq:DE}
\end{equation}
Here, the shift is given by
\begin{equation}
\Sigma_E(k)\equiv \sum_{k'\neq k}\frac{\hbar^2|g(k')|^2}{E-E_T-\frac{\hbar^2k'^2}{m}}+\frac{\hbar^2|g_b|^2}{E-E_T-E_b},
\label{eq:shift}
\end{equation}
where the sum arises from the coupling of the state $\ket{g_1}$ to the continuum and the last term arises from the coupling of $\ket{g_1}$ to a near threshold triplet bound state. Note that we will include this term in the model only when the background scattering length $a_{bg}$ is {\it large and positive}.

To convert the sums into integrals, we use $\sum_{k'\neq k}=[V/(2\pi)^3]\,P\,\int 4\pi k'^2dk'$, where $P$ denotes the principal part, and we define the volume-independent coupling strength
\begin{equation}
\hbar\tilde{g}(k')\equiv\sqrt{\frac{V}{(2\pi)^3}}\,\hbar\,g(k')\equiv V_{HF}\,\langle\tilde{k}'\ket{v_1},
\label{eq:couplingtwiddle}
\end{equation}
where $\langle\tilde{k}'\ket{v_1}$ is the spatial overlap integral of the vibrational wavefunction $\langle r|v_1\rangle$ with the continuum normalized momentum eigenstate. In the limit $r\rightarrow\infty$,
\begin{equation}
\langle r |\tilde{k'} \rangle=\frac{1}{\sqrt{(2\pi)^3}}\frac{\sin(k'r+\delta'_{bg})}{k'r},
\label{eq:twiddlek}
\end{equation}

To determine the shift, we define $E_b=-\hbar^2{\cal K}_b^2/m$, where ${\cal K}_b\equiv 1/a_{bg}$ for large $a_{bg}>0$ with $|a_{bg}|$ large compared to the range of the scattering potential. Using $E=E_T+\hbar^2k^2/m$,
and noting that $\sum_{k'\neq k}$ becomes the principal part $P$ of the integral over $k'$, we obtain
\begin{equation}
\Sigma_E(k)\equiv \frac{m|g_b|^2}{k^2+{\cal K}_b^2}+P\int_0^\infty 4\pi k'^2dk'\frac{m|\tilde{g}(k')|^2}{k^2-k'^2}.
\label{eq:shiftcontinuum}
\end{equation}

The asymptotic form of the scattering state, $\langle r|\Psi_E(t)\rangle$, in the limit $r\rightarrow \infty$, determines the s-wave phase shift. At large distance, the singlet and triplet molecular wavefunctions vanish. Hence, the scattering state is determined by the triplet continuum part of the wavefunction, $\langle r\rightarrow\infty|\Psi_E(t)\rangle\rightarrow \psi_T(r)\,\ket{T}\exp(-iEt/\hbar)$, where
\begin{equation}
\psi_T(r)=b_T(k)\langle r|k\rangle+\sum_{k'\neq k}b_T(k')\langle r|k'\rangle.
\label{eq:tripletscattering1}
\end{equation}
Using eqs.~\ref{eq:bTk'}~and~\ref{eq:b1k}, we obtain
\begin{equation}
b_T(k'\neq k)=-\frac{m\,g(k')}{k'^2-k^2}\frac{g^*(k)}{D(E)}\,b_T(k).
\end{equation}
With eq.~\ref{eq:tripletscattering1}, this yields
\begin{equation}
\hspace*{-0.125in}\psi_T(r)=b_T(k)\left\{\langle r|k\rangle-\sum_{k'\neq k}\frac{m\, g(k')}{k'^2-k^2}\frac{g^*(k)}{D(E)}\langle r|k'\rangle\right\}.
\label{eq:psiT1}
\end{equation}
From eqs.~\ref{eq:backgroundstate},~\ref{eq:couplingtwiddle}, and~\ref{eq:psiT1}, we obtain the scattering state as
\begin{eqnarray}
&\psi_T(r\rightarrow\infty)=\frac{b_T(k)}{\sqrt{V}}\frac{1}{kr}\Big\{\sin[kr+\delta_{bg}(k)]\nonumber\\
&-\frac{4\pi\, m\,k\,\tilde{g}^*(k)}{D(E)}\,P\int_0^\infty\frac{dk'\,k'\,\sin[k'r+\delta'_{bg}(k')]\,\tilde{g}(k')}{k'^2-k^2}\Big\}.
\label{eq:psiT2}
\end{eqnarray}

To evaluate the principal part appearing in eq.~\ref{eq:psiT2}, we note from eq.~\ref{eq:couplingtwiddle} that $\tilde{g}(k')$ is an even function of $k'$ since $\langle r|\tilde{k}'\rangle$, eq.~\ref{eq:twiddlek}, is even in $k'$.  Note that $\delta'_{bg}$ is an odd function of $k'$, i.e., the effective range expansion is $k'\cot[\delta'_{bg}(k')]=-1/a_{bg}+k'^2\,r_{bg}/2$ and the right side is even in $k'$. Hence the integrand is even in $k'$. Further, as shown below in the context of our simple model, $\tilde{g}(k')$ has only pure imaginary poles arising from the exponential form of the molecular bound state of size $R$. In the convergent half planes,  the sine function yields terms of the form $\exp(ik'r)\rightarrow \exp(-r/R)\rightarrow 0$, which make no contribution to the integral as $r\rightarrow\infty$. Hence, the principal part in the limit $r\rightarrow\infty$ is evaluated as

\begin{eqnarray}
&\frac{1}{2}\,P\int_{-\infty}^\infty\frac{dk'\,k'\,\sin[k'r+\delta'_{bg}(k')]\,\tilde{g}(k')}{k'^2-k^2}=\nonumber\\
&\hspace*{0.125in}\frac{\pi}{2}\,\tilde{g}(k)\,\cos[kr+\delta_{bg}(k)].
\label{eq:principal}
\end{eqnarray}

Using eq.~\ref{eq:principal} in eq.~\ref{eq:psiT2}, we obtain finally the asymptotic triplet scattering state,
\begin{widetext}
\begin{equation}
\psi_T(r\rightarrow\infty)=\frac{b_T(k)}{\sqrt{V}}\frac{1}{kr}\left\{\sin[kr+\delta_{bg}(k)]-\frac{2\pi^2\, m\,k\,|\tilde{g}(k)|^2}{D(E)}\,\cos[kr+\delta_{bg}(k)]\right\},
\label{eq:psiT3}
\end{equation}
\end{widetext}
where the input triplet continuum state is of energy $E=E_T+\hbar^2k^2/m$ and $E_T\equiv-a_{HF}/2-2\mu_B\,B$ is magnetic field dependent.

To determine the total phase shift $\Delta(k)\equiv\tilde{\Delta}(k)+\delta_{bg}(k)$, we write $\psi_T(r\rightarrow\infty)=A(k)\,\sin[kr+\tilde{\Delta}(k)+\delta_{bg}(k)]/(kr)$,
where $\tilde{\Delta}(k)$ is the resonant part of the phase shift and $\delta_{bg}(k)$ is the background part.
Then, comparing eq.~\ref{eq:psiT3} with $A(k)\cos\tilde{\Delta}\sin[kr+\delta_{bg}(k)]$ and $A(k)\sin\tilde{\Delta}\cos[kr+\delta_{bg}(k)]$, we obtain
\begin{equation}
\tan\tilde{\Delta}(k)=-\frac{2\pi^2 m\,k\,|\tilde{g}(k)|^2}{D(E)},
\label{eq:phaseopt}
\end{equation}
where the numerator determines the resonance width, which arises from the decay of the dressed molecular state into the continuum at a rate $\Gamma (k)$, where
\begin{equation}
\frac{\hbar\Gamma (k)}{2} = 2\pi^2m\,k|\tilde{g}(k)|^2,
\label{eq:decayrate}
\end{equation}
as is readily verified using Fermi's Golden rule.
\section{Zero Energy Scattering Length and Effective Range}
We determine the zero energy scattering length $a$ and the effective range $r_e$, from  the total phase shift $\Delta =\tilde{\Delta}+\delta_{bg}$  using
\begin{equation}
k\cot[\Delta(k)]=-\frac{1}{a}+\frac{k^2\,r_e}{2}.
\label{eq:effrangexpansion}
\end{equation}
With the elementary trigonometric relation between
$\cot(\tilde{\Delta}+\delta_{bg})$ and $\cot\tilde{\Delta}$ and $\cot\delta_{bg}$, we have
\begin{equation}
k\cot\Delta=\frac{(k\cot\tilde{\Delta})(k\cot\delta_{bg})-k^2}{k\cot\delta_{bg}+k\cot\tilde{\Delta}}.
\end{equation}
To expand eq.~\ref{eq:effrangexpansion} up to order $k^2$, we use
\begin{equation}
k\cot[\tilde{\Delta}(k)]=-\frac{D(E)}{2\pi^2 m\,|\tilde{g}(k)|^2}\equiv-\frac{1}{\tilde{a}}+\frac{k^2\,\tilde{r}_e}{2}.
\label{eq:tildeeffrangexpansion}
\end{equation}
and
\begin{equation}
k\cot[\delta_{bg}(k)]=-\frac{1}{a_{bg}}+\frac{k^2\,r_{bg}}{2},
\label{eq:bgexpansion}
\end{equation}
with obvious notation.

After some straightforward algebra, keeping terms up to order $k^2$ (we avoid the zero crossing, assuming $\tilde{a}+a_{bg}\neq 0$), we obtain the zero-energy scattering length
\begin{equation}
a=a_{bg}+\tilde{a}
\label{eq:scattlength}
\end{equation}
and the effective range,
\begin{equation}
r_e=\frac{\tilde{r}_e\tilde{a}^2+2\tilde{a}a_{bg}(\tilde{a}+a_{bg})+r_{bg}a_{bg}^2}{(\tilde{a}+a_{bg})^2}.
\label{eq:effrange1}
\end{equation}
Note that in the limit $\tilde{a}=0$, we have $r_e = r_{bg}$ and for $a_{bg}=0$, we have $r_e=\tilde{r}_e$ as expected.

We are interested in the zero energy scattering length and the effective range near resonance, where the resonant part of the zero energy scattering length is large in magnitude compared to that of the background scattering length. In the near resonance limit $|\tilde{a}|>>|a_{bg}|$, eq.~\ref{eq:effrange1} yields
\begin{equation}
r_e=\left[\tilde{r}_e+2\,a_{bg}+\frac{2\,|a_{bg}|^2}{\tilde{a}}\right]\left(1-\frac{2\,a_{bg}}{\tilde{a}}\right).
\label{eq:effrange2}
\end{equation}
Here, we ignore the effective range $r_{bg}$ for the background scattering states. For small $|a_{bg}|$, we can take $r_e=\tilde{r}_e$. In addition, we show in Appendix~\ref{simplemodel} using a simple model, that when  $|a_{bg}|$ is large compared to the singlet  molecular size, the leading contributions to $r_e$ from $\tilde{r}_e$ exactly cancel  the $a_{bg}$-dependent terms in the square bracket. Assuming a near resonance condition, where $|a_{bg}/\tilde{a}|$ is small, we can then  neglect the correction in the parenthesis.

In the following, we obtain both the zero-energy scattering length,  the optically-induced inelastic decay rate, and the effective range, including the modification arising from the optical fields. This is accomplished by expanding eq.~\ref{eq:tildeeffrangexpansion} up to order $k^2$. In this way, we obtain both the resonant part of the scattering length $\tilde{a}$ and corresponding effective range $\tilde{r}_e$.

For the resonant phase shift, we have using eqs.~\ref{eq:tildeeffrangexpansion} and~\ref{eq:DE} with $E=E_T+\hbar^2k^2/m$,
\begin{equation}
k\cot\tilde{\Delta}=-\frac{E_T-E_{g_1}-\Sigma_E(k)+\frac{\hbar^2k^2}{m}+D_{opt}(k)}{2\pi^2m|\tilde{g}(k)|^2},
\label{eq:CotTildeDelta}
\end{equation}
where  the optical contribution in the numerator is
\begin{equation}
D_{opt}(k)\equiv-\frac{\hbar|\Omega_1|^2}{4[\Delta_e+i\frac{\gamma_e}{2}+\frac{|\Omega_2|^2}{4\delta}]}.
\label{eq:DEOpt}
\end{equation}
Note that both $\Delta_e$ and $\delta$ are dependent on $E=E_T+\hbar^2k^2/m$ and are therefore $k^2$ dependent, altering the effective range.

Using $k\cot\tilde{\Delta}=-1/\tilde{a}+k^2\,r_e/2$, the resonant part of the zero energy scattering length is given by the $k=0$ contribution,
\begin{equation}
\frac{1}{\tilde{a}}=\frac{E_T-E_{g_1}-\Sigma_E(0)+D_{opt}(0)}{2\pi^2m|\tilde{g}(0)|^2}.
\label{eq:rescattlength}
\end{equation}
Since $E_T\equiv-a_{HF}/2-2\mu_B\,B$, we can write
\begin{equation}
E_T-E_{g_1}-\Sigma_E(0)\equiv -2\mu_B(B-B_\infty),
\label{eq:Btuning}
\end{equation}
where the resonance position $B_\infty$ includes the shift $\Sigma_E(0)$. We define the width $\Delta B$ of the resonance in terms of the background scattering length,  by
\begin{equation}
2\pi^2m|\tilde{g}(0)|^2\equiv |a_{bg}|\,2\mu_B\,\Delta B,
\label{eq:width}
\end{equation}
where  $\Delta B$ is positive by definition. We show in Appendix~\ref{simplemodel} that for a background scattering length that is large compared to the molecular size, the energy width $2\mu_B\,\Delta B$ and the shift $\Sigma_E(0)$ are equal in magnitude. Eq.~\ref{eq:rescattlength} determines how the optical fields control the zero energy scattering length $a=a_{bg}+\tilde{a}$.

In the absence of optical fields, $D_{opt}\rightarrow 0$, the resonant part of the scattering length is then $\tilde{a}[B]=-|a_{bg}|\Delta B/(B-B_\infty)$ and the zero energy scattering length takes the usual form~\cite{ChinFeshbach}
\begin{equation}
a[B]=a_{bg}-|a_{bg}|\frac{\Delta B}{B-B_\infty}.
\label{eq:ScattLengthNoOpt}
\end{equation}
We see that the zero crossing $a[B_0]=0$ occurs at a field $B_0$ below (above) resonance for $a_{bg}$ negative (positive).

The resonant part of the effective range is determined from the $k^2$ terms in the expansion of eq.~\ref{eq:CotTildeDelta},

\begin{eqnarray}
&\frac{k^2}{2}\tilde{r}_e=-\frac{\frac{\hbar^2k^2}{m}-k^2\left.\frac{\partial\Sigma_E(k)}{\partial (k^2)}\right|_{k=0}+k^2\left.\frac{\partial D_{opt}(k)}{\partial (k^2)}\right|_{k=0}}{2\pi^2m|\tilde{g}(0)|^2}\nonumber\\
& \hspace*{0.1in}+\frac{k^2\left.\frac{\partial |\tilde{g}(k)|^2}{\partial (k^2)}\right|_{k=0}[-2\mu_B(B-B_\infty)+D_{opt}(0)]}{2\pi^2m|\tilde{g}(0)|^4}.
\label{eq:8.5}
\end{eqnarray}

Eq.~\ref{eq:8.5} can be rewritten as
\begin{equation}
\tilde{r}_e=\tilde{r}_e^{(0)}+\tilde{r}_e'+\tilde{r}_{e}^{opt},
\label{eq:8.6}
\end{equation}
where $\tilde{r}_e^{(0)}=-2(\hbar^2/m)/(2\pi^2m|\tilde{g}(0)|^2)$  arises from the $\hbar^2k^2/m$ term in eq.~\ref{eq:8.5}, i.e.,  the relative kinetic energy. This term  is always present, i.e., even if the shift is independent of $k^2$. Using eq.~\ref{eq:width}, we obtain
\begin{equation}
\tilde{r}_e^{(0)}=-\frac{\hbar^2}{m\mu_B\Delta B\,|a_{bg}|},
\label{eq:8.8}
\end{equation}
which gives the resonant part of the effective range in the absence of optical fields when the {\it shift} $\Sigma_E(k)$  is independent of $k^2$. We see that broad resonances with large background scattering lengths, this contribution to the effective range will be small, while it can be large for narrow resonances with small background scattering lengths~\cite{HoNarrowFB,OHaraNarrowFB}.

The $\tilde{r}_e'$ term arises from the energy-dependent shift $\Sigma_E(k)$ and coupling $|\tilde{g}(k)|^2$, which may vary rapidly with $k^2$ when $|a_{bg}|$ is large, producing large contributions to the effective range. Using eq.~\ref{eq:rescattlength} for the resonant part of the scattering length $\tilde{a}$, this term can be written in the form
\begin{equation}
\tilde{r}_e'=\frac{2\left.\frac{\partial\Sigma_E(k)}{\partial (k^2)}\right|_{k=0}}{2\pi^2m|\tilde{g}(0)|^2}+\frac{2}{\tilde{a}|\tilde{g}(0)|^2}\left.\frac{\partial|\tilde{g}(k)|^2}{\partial (k^2)}\right|_{k=0}.
\label{eq:8.7}
\end{equation}
Finally, the optical fields alter the resonant part of the effective range,
\begin{equation}
\tilde{r}_e^{opt}=-\frac{2\frac{\partial D_{opt}(k)}{\partial (k^2)}|_{k=0}}{2\pi^2m|\tilde{g}(0)|^2}.
\label{eq:8.7opt}
\end{equation}

In Appendix~\ref{simplemodel}, we determine the optical field independent part of the effective range using a simple model for $|a_{bg}|>>R$, where $R$ is the effective size of the singlet vibrational state $|v_1\rangle$.
To evaluate $\tilde{r}_e'$, eq.~\ref{eq:8.7}, we use eq.~\ref{eq:10.3} to obtain
\begin{equation}
\left.\frac{\partial \Sigma_E(k)}{\partial (k^2)}\right|_{k=0}=-2\pi^2m|\tilde{g}(0)|^2|a_{bg}|\{2\,\theta[a_{bg}]-1\}.
\label{eq:11.3a}
\end{equation}
and eq.~\ref{eq:9.5}, which gives
\begin{equation}
\left.\frac{\partial |\tilde{g}(k)|^2}{\partial (k^2)}\right|_{k=0}=-|a_{bg}|^2|\tilde{g}(0)|^2.
\label{eq:11.3b}
\end{equation}
Then, eq.~\ref{eq:8.7} yields
\begin{eqnarray}
\tilde{r}_e'&=&-2\,|a_{bg}|\{2\,\theta[a_{bg}]-1\}-\frac{2\,|a_{bg}|^2}{\tilde{a}}\nonumber\\
&=&-2\,a_{bg}-\frac{2\,|a_{bg}|^2}{\tilde{a}},
\label{eq:11.3}
\end{eqnarray}
where the theta function assures that the first term is just $-2\,a_{bg}$, for either positive or negative $a_{bg}$.
We see from eq.~\ref{eq:11.3} that  $\tilde{r}_e'$ exactly cancels the corresponding terms in the square bracket of eq.~\ref{eq:effrange2}.

Hence, neglecting the small correction arising from the parenthesis in eq.~\ref{eq:effrange2}, the effective range for both small and large background scattering lengths takes the simple form
\begin{equation}
r_e=\tilde{r}_e^{(0)}+\tilde{r}_e^{opt},
\end{equation}
where $\tilde{r}_e^{opt}$ is given by eq.~\ref{eq:8.7opt} and $\tilde{r}_e^{(0)}$ is given by eq.~\ref{eq:8.8}. In the absence of optical fields, we see that $r_e=\tilde{r}_e^{0)}$, which is  usually obtained by ignoring the energy dependence of the shift and width.

As $D_{opt}(k)$ in eq.~\ref{eq:8.7opt} is a function of $E=E_T+\hbar^2k^2/m$, we can write $2\,\partial D_{opt}(k)/\partial (k^2)=2\,(\hbar^2/m)\partial D_{opt}(E)/\partial E$. Then, since $-2\,(\hbar^2/m)/(2\pi^2m|\tilde{g}(0)|^2)=\tilde{r}_e^{(0)}$, we have
\begin{equation}
r_e=\tilde{r}_e^{(0)}\left.\left[1+\frac{\partial D_{opt}(E)}{\partial E}\right|_{k=0}\,\right].
\label{eq:effrangeopt2}
\end{equation}
The $D_{opt}$ term enables optical control of the effective range, as it can be made to vary rapidly with energy near a dark-state resonance. In the following, we systematically examine the real and imaginary parts of eq.~\ref{eq:rescattlength} and eq.~\ref{eq:effrangeopt2}.

\section{Optical Control of the Scattering Length}

To find the real and imaginary parts of the zero energy ($k=0$) scattering length from eq.~\ref{eq:rescattlength}, we set $a=a_{bg}+\tilde{a}=a'+ia''$. For this purpose, it is convenient to define the magnetic field detuning,
\begin{equation}
\Delta_0=2\mu_B(B-B_\infty)/\hbar,
\label{eq:Delta0}
\end{equation} so that $-\hbar\Delta_0\equiv-2\mu_B\,(B-B_\infty)=E_T-E_{g_1}-\Sigma_E(0)$. Then, we obtain the simple form
\begin{equation}
a=a_{bg}-|a_{bg}|\beta\,\frac{1}{\Delta_0+\frac{|\Omega_1|^2}{4(\delta_e+i\gamma_e/2)}},
\label{eq:2.6}
\end{equation}
where $\beta\equiv 2\mu_B\,\Delta B/\hbar$. Here, we have defined
 \begin{equation}
 \delta_e=\Delta_e+\frac{|\Omega_2|^2}{4\delta},
 \label{eq:deltae}
 \end{equation}
Note that all the detunings are evaluated for $k=0$, i.e., with $E\rightarrow E_T=-a_{HF}/2-2\mu_B\,B$ the magnetic field dependent triplet energy.

The real part of the scattering length is then given by
\begin{equation}
a'=a_{bg}-|a_{bg}|\beta\,\frac{4\Delta_0\Gamma_2^2+|\Omega_1|^2\Gamma_2\delta+(\gamma_e\delta)^2\Delta_0}
{4(\Delta_0\Gamma_2+\delta|\Omega_1|^2/4)^2+(\gamma_e\delta\Delta_0)^2},
\label{eq:3.4}
\end{equation}
where $\Gamma_2\equiv\delta\,\delta_e= \delta\Delta_e+|\Omega_2|^2/4$ has a dimension of frequency {\it squared}.
Here, the one-photon detuning, eq.~\ref{eq:Deltae}, is $\Delta_e=\omega_1-(E_e-E_T)/\hbar$,   while eq.~\ref{eq:delta} defines the two-photon detuning $\delta=\omega_2-\omega_1-(E_T-E_{g_2})/\hbar$.

The corresponding imaginary part is
\begin{equation}
a''=-|a_{bg}|\,\frac{\beta}{2}\,\frac{\gamma_e|\Omega_1|^2\delta^2}
{4(\Delta_0\Gamma_2+\delta|\Omega_1|^2/4)^2+(\gamma_e\delta\Delta_0)^2}.
\label{eq:4.2}
\end{equation}
The imaginary part of the scattering rate causes inelastic loss, which arises from optical scattering,  with a two-body rate constant $K_2({\rm cm}^3/s)=-8\pi\hbar\,a''/m$ in the $k=0$ limit.\\

The dark state method offers many options for controlling  the  scattering length, the inelastic rate, the resonance width, and the effective range. These include varying the frequencies of the two optical fields, choosing the magnetic field detuning and controlling the amplitudes of the optical fields in space and time. For the initial discussion, we reproduce here the figures from our paper~\cite{WuOptControl}.\\

\begin{figure}[htb]
\begin{center}
\includegraphics[width=3.0in]{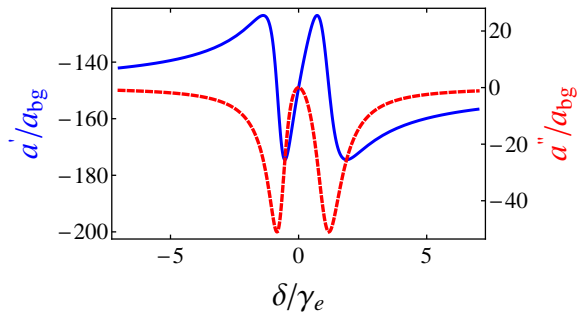}
  \caption{(color online). Scattering length as a function of the effective two-photon detuning  $\delta$ in units of $\gamma_{e}$. Real $a^{'}/a_{bg}$ (Top blue curve) and imaginary $a^{''}/a_{bg}$ (Bottom dashed red curve). From Ref.~\cite{WuOptControl}. }
\label{fig:fig2}
\end{center}
\end{figure}

Fig.~\ref{fig:fig2} shows the real and imaginary parts of the scattering length  as a function of the two-photon detuning $\delta$. We use the parameters  for $^6$Li: $\Delta B=300$ G,  $2\mu_B/\hbar=2\pi\times 2.8$ MHz/G, $ \gamma_{e}=2\pi \times 11.8$ MHz, and $a_{bg}=-1405\,a_0$; We take $\Omega_{1}=0.8 \gamma_{e}$, $\Omega_{2}=2 \gamma_{e}$, $\omega_2=\omega_{eg_{2}}$, $B-B_{0}= 2$ G.

\begin{figure}
\begin{center}
\includegraphics[width=3.0 in]{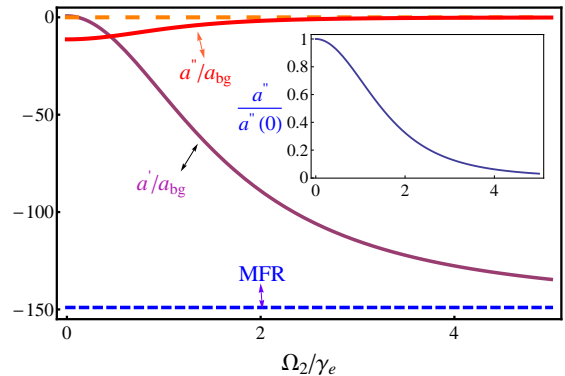}
\caption[example]
   { \label{fig:fig3}
(color online). Real $a^{'}/a_{bg}$ and imaginary $a^{''}/a_{bg}$ components of the scattering length as a function of  $\Omega_{2}/\gamma_{e}$ for $\Omega_{1}= 5\,\gamma_{e}$, and $\delta=0.05\,\gamma_{e}$. All other parameters are the same as in Fig.~\ref{fig:fig2}: The dashed blue line at the bottom is the scattering length without the laser fields (magnetic Feshbach resonance);  The dashed orange line at the top denotes $a''=0$. Inset: Loss ratio between the ``dark-state'' scheme and a typical single laser scheme (where $\Omega_{2}=0$) as a function of $\Omega_{2}/\gamma_{e}$. From Ref.~\cite{WuOptControl}. }
\end{center}
\end{figure}

\begin{figure}
\begin{center}
\includegraphics[width=3.0 in]{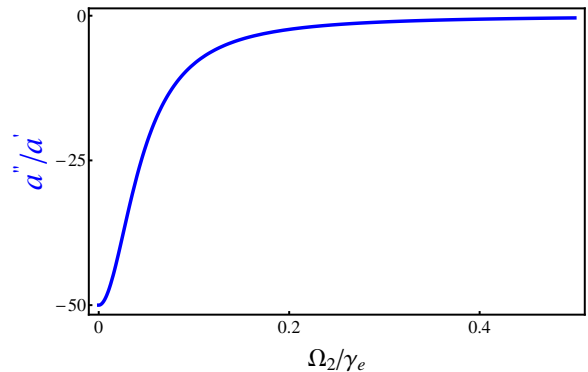}
\caption{ \label{fig:fig4}
(color online). The ratio of the imaginary to the real part of the scattering length  $a^{''}/a^{'}$  as a function of  $\Omega_{2}/\gamma_{e}$ for $\Omega_{1}= 5\,\gamma_{e}$, and $\delta=0.05\,\gamma_{e}$. All other parameters are the same as in Fig.~\ref{fig:fig3}.  }
\end{center}
\end{figure}

The dark-state optical control method enables the suppression of spontaneous scatter, which would cause
substantial loss and heating if the $\Omega_1$ beam were applied alone, as in Ref.~\cite{RempeOptControl}. Analogous to dark state methods for controlling the ratio of absorption to dispersion, we can control the ratio $a''/a'$. Assuming that the resonant part of the scattering length $a'$ is large compared to $a_{bg}$, $a''/a'$ is given by
\begin{equation}
\frac{a''}{a'}=-\frac{1}{2}\frac{\gamma_e|\Omega_1|^2\delta^2}
{\Delta_0\,[4\Gamma_2^2+(\gamma_e\delta)^2]+|\Omega_1|^2\Gamma_2\delta},
\label{eq:5.1}
\end{equation}
where we recall that $\Gamma_2\equiv \delta\Delta_e+|\Omega_2|^2/4$.
For $\Delta_0\neq0$, we see that  loss is suppressed compared to elastic scattering by the square of the two-photon detuning $\delta$. For $\Delta_0=0$, and large $\Omega_2$, the ratio is $-2\gamma_e\delta/|\Omega_2|^2$, which can be made small for sufficiently large Rabi frequency, $\Omega_2$. Fig.~\ref{fig:fig3} shows the scattering length as a function of $\Omega_2$, demonstrating the suppression of $a''$ as $\Omega_2$ is increased. The corresponding ratio $a''/a'$ is shown in Fig.~\ref{fig:fig4}.

In addition to inelastic loss arising from optical scattering, there is a small but finite photoassociation rate that arises from transitions between the triplet ground state and the excited singlet state. The triplet to triplet photoassociation transition is far away from resonance. For example, choosing the $^6Li$ excited singlet vibrational state $v'=70$, the closest vibrational state for the triplet state is $v'=62$, which is about $40$ GHz away from the singlet transition.  We will not discuss photoassociation here. However, we have shown theoretically that the small triplet to singlet photoassociation rate is also suppressed near the dark-state resonance.

\begin{figure}[htb]
\begin{center}
\includegraphics[width=3.0in]{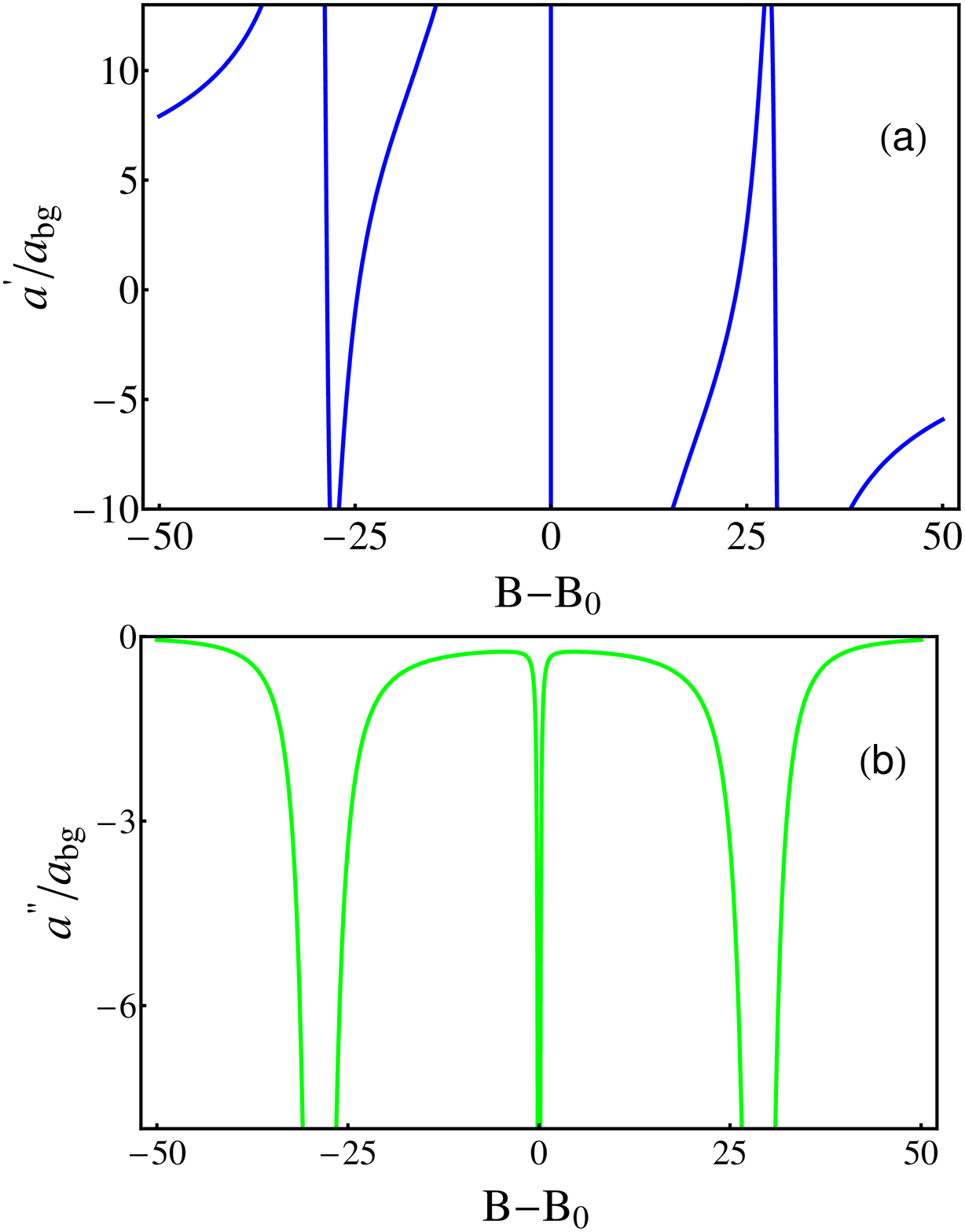}
\caption[example]
   { \label{fig:fig5}
(color online). Scattering length as a function of  $B-B_{0}$ for fixed laser parameters $\Omega_{1}=8\gamma_{e}$,
$\Omega_{2}=12\gamma_{e}$, $\omega_1=\omega_{eg_1}$, $\omega_2=\omega_{eg_2}$. All other parameters are the same as in Fig. 2: (a) $a^{'}/a_{bg}$ (b) $a^{''}/a_{bg}$. From Ref.~\cite{WuOptControl}. }
\end{center}
\end{figure}

Finally, the dark-state method produces ``artificial" narrow Feshbach resonances, which enable rapid changes in the scattering length for small changes in the magnetic field. Fig.~\ref{fig:fig5} shows the results for reasonably large value of $\Omega_2$, to clearly separate the broad and narrow resonances, which have the usual three-peak structure.

\section{Optical Control of the Effective Range}

In addition to controlling the real and imaginary parts of the scattering length, the dark-state method creates broad and narrow resonances, as shown in Fig.~\ref{fig:fig5}, which have different {\it optically controllable} effective ranges. The dark state method can be applied to both the broad  Feshbach resonance in $^6$Li, at 834 G, and the narrow resonance at 543 G. Using both of these, we can explore the role of the effective range over a wide range. As discussed below in more detail, for fixed two-photon detuning $\delta =0$, i.e., at the dark-state resonance, the scattering length remains at the preselected magnetic field value and the inelastic scattering length vanishes, while the effective range is dependent on the ratio of the intensities of the two optical beams. In this case,  the effective range can be varied at fixed scattering length, with negligible scattering.

We determine the optically-controlled part of the effective range, $\tilde{r}_e$ from  eq.~\ref{eq:effrangeopt2}, using eq.~\ref{eq:DEOpt}. In this case, we note that the one-photon detuning $\Delta_e$ defined by eq.~\ref{eq:Deltae}  and the two-photon detuning defined  by eq.~\ref{eq:delta} have a simple $E$-dependence, yielding $\partial\Delta_e/\partial E=1/\hbar$ and $\partial\delta/\partial E=-1/\hbar$. Hence,
\begin{eqnarray*}&\left.\partial [\Delta_e+i\gamma_e/2+|\Omega_2|^2/(4\delta)] /\partial E\right|_{k=0}\\
&=(1/\hbar)\left[1+|\Omega_2|^2/(4\delta^2)\right]
\end{eqnarray*}
which gives
\begin{equation}
r_e=\tilde{r}_e^{(0)}\left[1+\frac{|\Omega_1|^2\left[1+\frac{|\Omega_2|^2}{4\delta^2}\right]}
{4(\delta_e+i\gamma_e/2)^2}\right],
\label{eq:effrangeopt3}
\end{equation}
where $\delta_e$ is defined by eq.~\ref{eq:deltae} with all detunings evaluated for $k=0$, i.e, $E=E_T$.
Taking the real $r_e'$ and imaginary $r_e''$ parts of eq.~\ref{eq:effrangeopt3}, we obtain
\begin{equation}
r_e'=\tilde{r}_e^{(0)}\left\{1+
\frac{|\Omega_1|^2(4\delta^2+|\Omega_2|^2)[4\Gamma_2^2-(\gamma_e\delta)^2]}
{4[4\Gamma_2^2+(\gamma_e\delta)^2]^2}\right\}.
\label{eq:effrangereal}
\end{equation}

\begin{equation}
r_e''=-\tilde{r}_e^{(0)}\frac{|\Omega_1|^2(4\delta^2+|\Omega_2|^2)\Gamma_2\gamma_e\delta}
{[4\Gamma_2^2+(\gamma_e\delta)^2]^2},
\label{eq:effrangeimaginary}
\end{equation}
where $\Gamma_2 = \delta\delta_e=\delta\Delta_e+|\Omega_2|^2/4$.

From eq.~\ref{eq:8.8}, we recall that $r^{(0)}_e=-\hbar^2/(m\mu_B\Delta B|a_{bg}|)$. For the broad resonance in $^6$Li at 834 G, where $a_{bg}=-1405\,a_0$, $\Delta B=300$ G~\cite{BartensteinFeshbach}, we have $|r^{(0)}_e|\simeq 1\,a_0<<|a_{bg}|$. However, for the narrow Feshbach resonance at 543 G~\cite{zerocross,OHaraNarrowFB}, where $\Delta B\simeq 0.1$ G~\cite{OHaraNarrowFB} and $a_{bg}=62\,a_0$, $r^{(0)}_e\simeq -7\times 10^4\,a_0$, as noted in Ref.~\cite{OHaraNarrowFB}. Using the dark-state method,  the effective range can be widely varied.

In general, the one-field method $|\Omega_2|\rightarrow 0$ will alter the effective range as well as the scattering length. Assuming $|\Omega_2|^2/4<<\Delta_e\delta$,  and taking the one-photon detuning $\Delta_e=0$ for simplicity, we find $r_e''=0$ and $r_e'=\tilde{r}_e^{(0)}(1-|\Omega_1|^2/\gamma_e^2)$. However, $a''\neq 0$, which can cause substantial losses for a one-field method.

For the two-field method, the imaginary part $r_e''$ vanishes when either $\delta =0$ or $\Gamma_2=\delta\Delta_e+|\Omega_2|^2/4=0$. The former corresponds to the narrow peak at the center of Fig.~\ref{fig:fig5}, while the latter corresponds to the two side peaks. For the limiting case with $\Omega_2\neq 0$ satisfying $|\Omega_2|^2/4>>\delta\Delta_e$, and taking $\delta \rightarrow 0$, we obtain
\begin{equation}
\tilde{r}_e'=r^{(0)}_e\left[1+\frac{|\Omega_1|^2}{|\Omega_2|^2}\right],
\label{eq:case1}
\end{equation}
which shows that the effective range is negative, since $r^{(0)}_e<0$ from eq.~\ref{eq:8.8}, and increases in magnitude when the ratio of the Rabi frequencies for the two transitions $|\Omega_1/\Omega_2|>1$. Note that when $\delta \rightarrow 0$ for dark-state scheme, the loss is negligible and the scattering length does not change. The optical fields only modify the effective range.

\begin{figure}[htb]
\begin{center}
\includegraphics[width=3.0in]{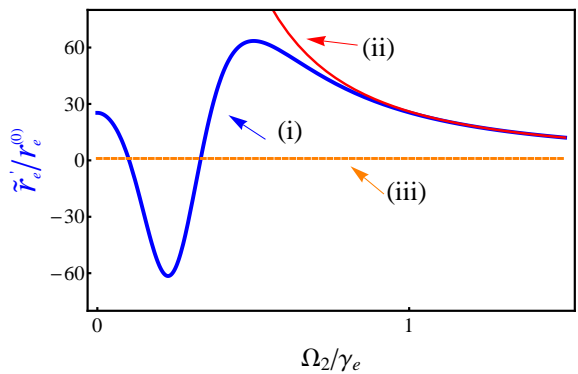}
  \caption{(color online). The real part of the optically controllable effective range ($\tilde{r}_e'$) in units of the effective range without optical fields ($r^{(0)}_e$) as a function of Rabi frequency $\Omega_2$ in units of $\gamma_{e}$. (i)~(Blue oscillatory curve) $\tilde{r}_e'/r^{(0)}_e$ for small two-photon detuning $\delta=0.05\,\gamma_e$.
  (ii)~(Red asymptotic curve) $\tilde{r}_e'/r^{(0)}_e$ for $\delta \rightarrow 0$. (iii) (Dashed orange curve) $\tilde{r}_e'/r^{(0)}_e=1$, i.e., the effective range in the absence of optical fields. }
\label{fig:effrange}
\end{center}
\end{figure}
The effective range as a function of $\Omega_{2}$ is shown in Fig.~\ref{fig:effrange} when two-photon detuning is small, $\delta/\gamma_e=0.05$. It is clear that the effective range can be widely modified by the laser fields.

For the case $\Gamma_2=0$, i.e., $\Delta_e=-|\Omega_2|^2/(4\delta)$, we obtain instead,
\begin{equation}
\tilde{r}_e'=r^{(0)}_e\left[1-\frac{|\Omega_1|^2(4\delta^2+|\Omega_2|^2)}{(2\gamma_e\delta)^2}\right],
\label{eq:case2}
\end{equation}
In this case, for small two-photon detuning $\delta$ and large one photon detuning $\Delta_e$, the effective range is large and positive, since $r^{(0)}_e<0$.

\section{Application to $^6$Li Magnetic Feshbach Resonances}

 In  $^6$Li, $\ket{g_1}$ is the $v=38$ Feshbach resonance state. We can take  $\ket{g_2}$ to be another lower lying vibrational state,  which are essentially uncoupled to the triplet state, since the nearest state $v=37$ is lower in energy than the $v=38$ state by  55.8 GHz, while the $v=36$ state is lower by 289 GHz~\cite{LiVibStates}.  We  employ the bound-to-bound singlet molecular transition from the ground $^1\Sigma^+_g\,(N=0)$ state to the excited $A^1\Sigma^+_u\, (N=1)$ state. Starting from the $v=38$ Feshbach resonance state,  the best Franck-Condon factor~\cite{CoteJMS1999} arises in a transition to the $v'=70$ vibrational state, which we take as $\ket{e}$. The nominal wavelength for this transition is 672.66 nm, which is readily accessible  with a  diode laser. A second diode laser excites the $v=37$ or $v=36$ to $v'=70$ transition.

We determine the Rabi frequencies from the known dipole transition matrix element. For the $v=38\rightarrow v'=70$ transition, the oscillator strength is $f_{eg}=0.035$~\cite{CoteJMS1999}, where
$f_{eg}=\mu_0'^2/[3(\,e\,a_{eg})^2]$. Here, $\mu_0'$ is the $z'$ component (along the internuclear axis)
of the electronic transition dipole moment between the selected
vibrational states,  $e$ is the electron charge, and
$a_{eg}=\sqrt{\hbar/(2m_e\omega_{eg})}$ is the electron harmonic
oscillator length scale for an electronic transition of frequency
$\omega_{eg}$. At $\lambda = 672.66$ nm, $e\,a_{eg}=6.9$ D (1 D =
1 Debye $\equiv 10^{-18}$ esu-cm.) Then $\mu_0'= 2.2$ D. The
corresponding laboratory dipole operator is $\mu^{(1)}_q=\mu_0'\,{\cal D}^{(1)}_{0\,q}(\theta,\varphi)$, where $\theta,\varphi$ are the Euler angles of the molecular internuclear axis with respect to the laboratory frame.

For a $\Delta M=0,\pm 1$ transition in the laboratory frame from
the $N=0$ $v=38$ ($J=0$) ground state to the $N=1$ $v=70$
($J=1,M$) excited state, the transition matrix elements are then
all $\mu\equiv\mu_0'/\sqrt{3}=1.3$ D. The corresponding Rabi
frequency is $\Omega_R({\rm Hz}) = 4.37 \,{\rm MHz}\,\mu({\rm
D})\sqrt{I({\rm mW/mm}^2)}$, which yields
$\Omega_1=5.7\, {\rm MHz}\,\sqrt{I({\rm mW/mm}^2)},$
where $I$ is the laser intensity.
This result is nearly identical to that given in
Ref.~\cite{HuletMolecProbe} for the $v'=68$ excited state.

For the trapped atoms, the applied fields will produce an effective light-shift potential, $U_{LS}=|\Omega_R'|^2/(4 \Delta_{laser})$ as well as spontaneous scattering at a rate $|\Omega_R'|^2/(4 \Delta_{laser}^2\,\tau_{spont})$, where $\Omega_R'$ is the Rabi frequency for the atomic D2 transition.
For $^6$Li, $\tau_{spont}=27$ ns and the molecular transition is $\simeq 1.66$ nm ($\simeq -1.1$
THz) red detuned from the free atom transition at 671 nm. The atomic
transition dipole moment is 5.9 D, so that $\Omega_R'=(5.9/1.3)\times\Omega_1=25.8\, {\rm MHz}\,\sqrt{I({\rm mW/mm}^2)}$. For $\Omega_1=10\,\gamma_e = 120$ MHz, $\Omega_R'=540$ MHz, the corresponding free atom
scattering rate is only $2.2$/s, which is negligible for the
proposed experiments, where the hold time in the weakly
interacting regime will $<<1$s.

For these parameters, the corresponding attractive light-shift potential,  $U_{LS}=66$ kHz, or $U_{LS}\simeq k_B\times 3\,\mu$K. This can be eliminated when necessary using a repulsive potential, provided by a spatially-matched blue-detuned beam at 532 nm.

\section{Conclusions}

The molecular dark-state method permits control of two-body scattering parameters, while suppressing light-induced inelastic loss and heating compared to single-field control techniques. As narrow features are produced by these dark-state quantum interference methods, optical fields can induce a strong dependence of the scattering phase shifts on the relative kinetic energy of the colliding atoms, enabling control of the effective range as well as the zero-energy scattering amplitude.


\appendix
\section{A Simple Model}
\label{simplemodel}

When the background scattering length $a_{bg}$ is large in magnitude compared to the molecular size, as it is for the broad resonance in $^6$Li, it is instructive to evaluate the width given by eq.~\ref{eq:width} and the shift  $\Sigma_E(k)$ given by eq.~\ref{eq:shiftcontinuum}, using a simple model.

 We assume that the resonant singlet molecular state $|v_1\rangle$ and the triplet bound state $|b\rangle$  have the simple forms
\begin{eqnarray}
\langle r\ket{v_1}&=&\frac{1}{\sqrt{2\pi R}}\frac{e^{-r/R}}{r}\nonumber\\
\langle r\ket{b}&=&\frac{1}{\sqrt{2\pi a_{bg}}}\frac{e^{-r/a_{bg}}}{r},
\label{eq:9.1}
\end{eqnarray}
where we include the $\ket{b}$ triplet state only for {\it large positive} scattering lengths, where $a_{bg}>>R$. Otherwise, this state is omitted from the calculation. Hence, there is an implied $\theta[a_{bg}]$ unit step function in the following. We assume that the background scattering states are {\it everywhere} given by  eq.~\ref{eq:twiddlek}, i.e., we ignore the small region of rapid oscillation in the deep part of the triplet potential well. Then, we easily obtain
\begin{equation}
\langle\tilde{k}\ket{v_1}=\frac{R^{3/2}}{\pi}\frac{1}{\sqrt{1+k^2a_{bg}^2}}\frac{(1-a_{bg}/R)}{1+k^2R^2}.
\label{eq:9.2}
\end{equation}
From eq.~\ref{eq:9.1}, we also have
\begin{equation}
\langle b\ket{v_1}=\frac{2\sqrt{a_{bg}R}}{R+a_{bg}},
\label{eq:9.3}
\end{equation}
where we assume a weakly bound (near threshold) triplet state of energy $E_b=-\hbar^2/(ma_{bg}^2)$, which arises for large positive (background) scattering lengths. Here, we again neglect the small region in the deep part of the molecular potentials, where the overlap integral of the singlet and triplet molecular states oscillates rapidly.
For $|a_{bg}|>>R$, we then have from eq.~\ref{eq:couplingtwiddle}~and~\ref{eq:9.2},
\begin{equation}
|\hbar\tilde{g}(k)|^2=|V_{HF}|^2\,\frac{R|a_{bg}|^2}{\pi^2}\frac{1}{1+(ka_{bg})^2}.
\label{eq:9.5}
\end{equation}
while eq.~\ref{eq:coupling}~and~\ref{eq:9.3} give for positive $a_{bg}>>R$,
\begin{equation}
|\hbar g_b|^2=|V_{HF}|^2\,\frac{4R}{a_{bg}}=|V_{HF}|^2\,\frac{4R}{|a_{bg}|}.
\label{eq:9.6}
\end{equation}
 Using eq.~\ref{eq:9.5}, the principal part term in eq.~\ref{eq:shiftcontinuum} is readily shown to be $-2\pi^2m|\tilde{g}(k)|^2/|a_{bg}|$. Then, we have
 \begin{equation}
 \Sigma_E(k)=\frac{m|g_b|^2|a_{bg}|^2}{1+(ka_{bg})^2}\,\theta[a_{bg}]-\frac{2\pi^2m|\tilde{g}(0)|^2}{|a_{bg}|[1+(ka_{bg})^2]},
 \label{eq:10.2}
 \end{equation}
where we include a unit step function to indicate that the contribution from the triplet bound state is to be used only when the background scattering length is large and positive.
Now, eqs.~\ref{eq:9.5}~and~\ref{eq:9.6} show that $m|g_b|^2|a_{bg}|^2=2\times 2\pi^2m|\tilde{g}(0)|^2/|a_{bg}|$. Hence,
\begin{equation}
\Sigma_E(k)=\frac{2\pi^2m|\tilde{g}(0)|^2}{|a_{bg}|[1+(ka_{bg})^2]}\{2\,\theta[a_{bg}]-1\}.
\label{eq:10.3}
\end{equation}

 From eq.~\ref{eq:10.3}~and~\ref{eq:width}, we see that for $|a_{bg}|>>R$, the width $\Delta B$ and the shift $\Sigma_E(0)$ are related by
\begin{equation}
\Sigma_E(0)=2\mu_B\Delta B\,\{2\,\theta[a_{bg}]-1\}.
\label{eq:11.2}
\end{equation}
Hence, the magnitude of the shift is equal to the twice the width $\mu_B\Delta B$.

We can apply this simple model to the three broad resonances in $^6$Li for 1-2, 1-3, and 2-3 mixtures of the three lowest hyperfine states, which are described in Table~\ref{Feshbach1} and Table~\ref{Feshbach2}.  The measured parameters for the $^6$Li Feshbach resonances are given in Table~\ref{Feshbach3}, taken from ref.~\cite{BartensteinFeshbach} for the broad resonances and from ref.~\cite{OHaraNarrowFB} for the narrow resonance. Recently, improved $^6$Li Feshbach resonance parameters have been obtained by using radio-frequency spectra of dimer pairs in very low density samples, which enables resolution of individual trap-radial-vibrational states~\cite{ZurnJochimFB}.
\begin{table}
\caption{Dominant singlet and triplet molecular states in the molecular (interior) basis $|S\,M_s;I\,M_I\rangle$ for $^6$Li Feshbach resonances. $M$ is the total magnetic quantum number for a pair of colliding atoms.\label{Feshbach1}}
\vspace*{0.25in}
\begin{tabular}{|c|c|c|c|}
             \hline
             Mixture&M &Singlet & Triplet \\
             \hline
             1-2 (B) &0& $(2\sqrt{2}|00;00\rangle-|00;20\rangle)/3$ & $|1-1;11\rangle$   \\
             1-2 (N) &0& $(|00;00\rangle+2\sqrt{2}|00;20\rangle)/3$& $|1-1;11\rangle$\\
             1-3    &-1& $|00;2-1\rangle$& $|1-1;10\rangle$ \\
             2-3    &-2& $|00;2-2\rangle$ & $|1-1;1-1\rangle$ \\
             \hline
           \end{tabular}
           \end{table}
                      \begin{table}
\caption{Triplet energy $E_T$ and singlet-triplet coupling $V_{HF}$ for Feshbach resonances in $^6$Li. $V_{HF}$  arises from the effective hyperfine interaction $a_{HF}(\mathbf{I}_1\cdot\mathbf{S}_1+\mathbf{I}_2\cdot\mathbf{S}_2)$, where $a_{HF}/h=152.1$ MHz. $E_T$ is the Zeeman-hyperfine energy for the given triplet molecular state, with $\mu_B$ the Bohr magneton $\mu_B/h=1.4$ MHz/G.  For the narrow (N) Feshbach resonance in the 1-2 mixture, the coupling is second order in the hyperfine interaction and $|10;10\rangle$ is the dominant off-resonant intermediate state. \label{Feshbach2}}
\vspace*{0.25in}
\begin{tabular}{|c|c|c|}
             \hline
             Mixture&$V_{HF}$ (MHz)&$E_T$ \\
             \hline
             1-2 (B)&$-3\,a_{HF}/(2\sqrt{3})=-131.6$ &$-2\mu_B\,B-a_{HF}/2$ \\
             1-2 (N)&$-a_{HF}^2/(E_{g_1}\sqrt{6})=-5.9$& $-2\mu_B\,B-a_{HF}/2$\\
             1-3 & $a_{HF}/2=76.0$ &$-2\mu_B\,B$ \\
             2-3& $a_{HF}/\sqrt{2}=107.5$ &$-2\mu_B\,B+a_{HF}/2$ \\
             \hline
           \end{tabular}
           \end{table}

\begin{table}
\caption{Feshbach resonance parameters for binary mixtures of the three lowest hyperfine states in $^6$Li.  The broad (B) resonance location $B_\infty$, width $\Delta B$, and background scattering length $a_{bg}$ in bohr ($a_0$) are taken from ref.~\cite{BartensteinFeshbach}. The narrow (N) Feshbach resonance parameters are taken from ref.~\cite{OHaraNarrowFB}.\label{Feshbach3}}
\vspace*{0.25in}
\begin{tabular}{|c|c|c|c|}
             \hline
             Mixture&$B_\infty$(G)& $\Delta B$& $a_{bg}(a_0)$  \\
             \hline
             1-2 (B) & 834 & 300 &-1405  \\
             1-2 (N) & 543 & 0.1 & +62\\
             1-3  & 690 & 122 &-1727 \\
             2-3  & 811 & 222 &-1490\\
             \hline
           \end{tabular}
\end{table}

   From eq.~\ref{eq:11.2}, for $a_{bg}<0$, we have $\Sigma_E(0)=-2\mu_B\Delta B$. This result can be used to estimate the energy $E_{g_1}$ of the resonant singlet bound state from the locations and widths of the broad resonances given in Table~\ref{Feshbach3}.  At resonance we have $E_T(B_\infty)=E_{g_1}+\Sigma_E(0)$. Then,
   \begin{equation}
   E_{g_1}=E_T(B_\infty)-\Sigma_E(0)\simeq E_T(B_\infty)+2\,\mu_B\Delta B
   \label{eq:energyv1}
   \end{equation}
   should be the same for all of the broad resonances.  For the narrow resonance, we assume that the shift $\Sigma_E(0)\simeq 0$.

We can also compute the effective size $R$ of the $\ket{v_1}$ state, eq.~\ref{eq:9.1}. Using eq.~\ref{eq:width}, eq.~\ref{eq:9.5}, and eq.~\ref{eq:couplingtwiddle}, we obtain
\begin{equation}
R=\frac{\hbar^2\,\mu_B\,\Delta B}{m\,|a_{bg}||V_{HF}|^2}.
\label{eq:effsize}
\end{equation}
Expressing $R$ in bohr units $a_0$, we have, $R/a_0=8.4\times 10^5\,\Delta B(G)/[|a_{bg}(a_0)||V_{HF}(\mbox{MHz})|^2]$, which also should be the same for all of the broad resonances.

Using the parameters in Table~\ref{Feshbach2} and Table~\ref{Feshbach3} and $a_{HF}/(4\mu_B)=27\,G$, we  obtain from eq.~\ref{eq:energyv1} and eq.~\ref{eq:effsize} the results given in Table~\ref{Feshbach4}.
\begin{table}
\caption{Estimated singlet vibrational energy $E_{g_1}$ and size $R$ for Feshbach resonances in $^6$Li.\label{Feshbach4}}
\vspace*{0.25in}
\begin{tabular}{|c|c|c|c|}
             \hline
             Mixture&$E_{g_1}/h$ (GHz)& $R(a_0)$  \\
             \hline
             1-2 (B) & $-2\mu_B(561\,G)/h=-1.57$&10.3 \\
             1-3  & $-2\mu_B\,(568\,G)/h=- 1.59$ & 10.2 \\
             2-3  & $-2\mu_B(562\,G)/h=-1.57$ & 10.8\\
             1-2 (N)& $-2\mu_B (570\,G)/h=-1.60$& \\
             \hline
           \end{tabular}
\end{table}

 The singlet energies obtained from all four resonances are nearly identical.  Hence, the approximation $\Sigma_E(0)=-2\mu_B\Delta B$ appears to be reasonably accurate for the large negative background scattering lengths in $^6$Li, which are large in magnitude compared to the  size of the resonant molecular state. Further, the nearly constant value of $R$ validates the scaling given by eq.~\ref{eq:effsize} based on the simple model.

We can compare the results obtained for $R$ with that expected using the overlap integrals for the true wavefunctions in the singlet and triplet potentials. Using eq.~\ref{eq:9.5} and eq.~\ref{eq:couplingtwiddle}, we  define an effective size $R\rightarrow R_{eff}=(\pi^2/|a_{bg}|^2)|\langle\tilde{k}|v_1\rangle|^2_{k\rightarrow 0}$. Then, with $\psi_{v_1}(r)\equiv u_{v_1}/r$ and $\psi_{Tk}(r)\equiv u_{Tk}/r$,
\begin{equation}
R_{eff}\equiv\frac{2\pi}{(ka_{bg})^2}\left|\int_0^\infty dr\,u_{v_1}(r)u_{Tk}(r)\right|^2_{k\rightarrow 0},
\end{equation}
where we take the triplet scattering state to be normalized so that $u_{Tk}(r\rightarrow\infty)=\sin[k(r-a_{bg})]$ as $k\rightarrow 0$. For the simple model,  eq.~\ref{eq:9.1}, we have $u_{v_1}(r)=\exp(-r/R)/(\sqrt{2\pi R})$. Taking $u_T(r)=\sin[k(r-a_{bg})]$ everywhere and assuming $|a_{bg}|>>R$, we immediately obtain $R_{eff}=R$ as assumed above. We have determined the overlap integral of the states obtained for the real triplet and singlet potentials (which yield the correct highest bound states), using the above normalization for $u_T(r)$ and the triplet scattering length obtained from the triplet scattering state $a_{bg}=a_T\simeq -2046\,a_0$. This yields yield $R_{eff}\simeq 11.4\,a_0$, within 10\% of the value $R\simeq 10.5\,a_0$ obtained from the broad Feshbach resonance parameters.

\end{document}